\documentclass[preprint,showpacs,preprintnumbers,amsmath,amssymb]{revtex4}

\usepackage{graphicx}
\usepackage{dcolumn}
\usepackage{bm}
\usepackage{latexsym}
\usepackage{amsfonts}
\usepackage{amssymb}
\usepackage{amsmath}
\usepackage{color}

\begin{document}

\preprint{KUNS-2191, DCPT-09/11}

\title{
Inflationary Universe with Anisotropic Hair
}

\author{Masa-aki Watanabe$^{1)}$}
\author{Sugumi Kanno$^{2)}$}
\author{Jiro Soda$^{1)}$}
\affiliation{1) Department of Physics,  Kyoto University, Kyoto, 606-8501, 
Japan}
\affiliation{2) Centre for Particle Theory, Department of Mathematical 
Sciences, Durham University, Science Laboratories, South Road, Durham, 
DH1 3LE, United Kingdom}

\date{\today}

\begin{abstract}
We study an inflationary scenario with a vector field coupled with
an inflaton field and show that the inflationary universe is endowed 
with anisotropy for a wide range of coupling functions.
This anisotropic inflation 
is a tracking solution where the energy 
density of the vector field follows that of the inflaton field
irrespective of initial conditions.

We find a universal relation between the anisotropy and
  a slow-roll parameter of inflation. 
Our finding has observational implications
and gives a counter example to the cosmic no-hair conjecture.
\end{abstract}

\pacs{98.80.Cq, 98.80.Hw}
\maketitle

\section{Introduction}

Recent developments of precision cosmology have yielded a slight 
shift of an inflationary paradigm~\cite{Komatsu:2008hk}. 
Before the precision cosmology, zeroth order predictions of
inflationary scenarios were sufficient.
Indeed, curvature fluctuations had been supposed to be 
statistically homogeneous, isotropic, gaussian and almost scale 
invariant. However, because of progress in observations, 
we are now forced to look at fine structures of fluctuations 
such as spectral tilt, non-gaussianity, parity violation, and 
so on~\cite{Baumann:2008aq}. 
 In fact, we need theoretical predictions at a percent level.
 Those precise predictions of inflationary scenarios
 will provide a clue to understand fundamental physics 
such as superstring theory when they are compared with 
observations.

In this paper, we focus on a role of a vector field
 in the early universe~\cite{Kanno:2006ty}.
Of course, no one doubts existence of vector fields. 
At the same time, it is widely believed vector hair 
will disappear during the inflation conforming to the
cosmic no-hair conjecture~\cite{Wald:1983ky}. 
However, recently, it is shown that anisotropic hair 
in the inflationary universe can exist~\cite{Golovnev:2008cf,Kanno:2008gn},
although there may be perturbative instability in this specific
realization~\cite{Himmetoglu:2008zp}.
Hence, it is worth seeking other models.  
 At this point, we should recall that primordial magnetic fields
are produced during inflation~\cite{Turner:1987bw}. For example, 
 the nonminimal kinetic term of vector fields in supergravity 
can be used to generate the primordial cosmological magnetic 
fields~\cite{Martin:2007ue}.
This fact suggests that
 we have a vector hair during inflation. 
 Here, there is a prejudice that the vector hair is negligibly small and
it is legitimate to ignore the backreaction of magnetic fields to geometry.
 However, in the context of the precision cosmology,
we should not neglect the backreaction
if it is around a percent level~\cite{Pullen:2007tu}.
Hence, it is important to quantify how small it is.
Based on this observation, 
we study an inflationary scenario
 where the inflaton is coupled with the kinetic term of
 a massless vector field. Apparently, our model is free from instability. 
Interestingly, we find a tracking behavior of the energy density of the vector 
field. As a consequence, we show that there exist sizable vector hair 
quite generally. That yields a percent level anisotropic inflation. 

It should be stressed that the presence of the vector hair in the early 
universe breaks the rotational invariance and therefore provides various 
interesting phenomenological consequences~\cite{Yokoyama:2008xw}. 
Moreover, anisotropic inflation might give rise to
 a percent level correlation between primordial gravitational waves
 and cosmic microwave background radiations (CMB), which might be testable 
by CMB observations near future~\cite{Kanno:2008gn}.
 Therefore, ``hairy inflation" is phenomenologically rich.

\section{Basic equations}
\label{sc:basic}

We consider the following action for the gravitational field, the inflaton
 field $\phi$ and the
vector field $A_\mu$ coupled with $\phi$:
\begin{eqnarray}
S&=&\int d^4x\sqrt{-g}\left[~\frac{1}{2\kappa^2}R
-\frac{1}{2}\left(\partial_\mu\phi\right)\left(\partial^{\mu}\phi\right)
-V(\phi)-\frac{1}{4} f^2 (\phi) F_{\mu\nu}F^{\mu\nu}  
~\right] \ ,
\label{action1}
\end{eqnarray}
where $g$ is the determinant of the metric, $R$ is the
Ricci scalar, $V(\phi)$ is the inflaton potential, $f(\phi)$ is the coupling function of the inflaton field to the vector one, respectively.
 The field strength of the vector field is defined by 
$F_{\mu\nu}=\partial_\mu A_\nu -\partial_\nu A_\mu$. 
Thanks to the gauge invariance, we can choose the gauge $A_0 =0$.
 Without loss of generality,
we can take $x$-axis in the direction of the vector.
Hence, we take the homogeneous fields of the form
$
A_\mu=(~0,~A_x(t),~0,~0~)
$
and 
$ 
\phi=\phi(t) \ .
$
Note that we have assumed the direction of the vector field does
not change in time, for simplicity. 
This field configuration holds the plane symmetry in the plane 
perpendicular to the vector.
Then, we take the metric to be 
\begin{eqnarray}
ds^2=- dt^2+e^{2\alpha(t)}\left[~ 
e^{-4\sigma(t)}dx^2    
+e^{2\sigma(t)}\left( dy^2 + dz^2\right)~\right] \ ,
\label{metric}
\end{eqnarray}
where the cosmic time $t$ is used.
Here, $e^\alpha$ is an isotropic scale factor and $\sigma$ represents
a deviation from the isotropy.  With above ansatz, 
one obtains the equation of motion for the vector field which is
easily solved as
\begin{eqnarray}
\dot{A_x} = f^{-2}(\phi ) e^{-\alpha -4\sigma}p_{A}, 
\label{eq:Ax}
\end{eqnarray}
where an overdot denotes the derivative with respect to the cosmic time $t$
and $p_A$ denotes a constant of integration. 
Substituting (\ref{eq:Ax}) into other equations, we obtain basic equations
\begin{eqnarray}
\dot{\alpha}^2  &=& \dot{\sigma}^2
+\frac{\kappa^2}{3}\left[ \frac{1}{2} \dot{\phi}^2+V(\phi)
+\frac{p_{A}^2}{2}f^{-2} (\phi) e^{-4\alpha-4\sigma }  \right] \ , 
\label{hamiltonian}\\
\ddot{\alpha} &=& -3\dot{\alpha}^2 + \kappa ^2 V(\phi )
 +\frac{\kappa ^2 p_{A}^2}{6}f^{-2}(\phi )e^{-4\alpha -4\sigma}, 
\label{evolution:alpha}\\
\ddot{\sigma} &=& -3\dot{\alpha}\dot{\sigma} 
+ \frac{\kappa ^2 p_{A}^2}{3}f^{-2}(\phi )e^{-4\alpha -4\sigma} 
\label{eq:sigma}, \\
\ddot{\phi} &=& -3\dot{\alpha}\dot{\phi} -V'(\phi ) 
+ p_{A}^2 f^{-3}(\phi )f'(\phi ) e^{-4\alpha -4\sigma } 
\label{eq:phi} \ ,
\end{eqnarray}
where a prime denotes the derivative with respect to $\phi$.

From Eq.(\ref{hamiltonian}), we see the effective potential
$
V_{\rm eff} = V + p_A^2 f^{-2} e^{-4\alpha -4\sigma}/2
$
determines the inflaton dynamics. As the second term is
coming from the vector contribution, we refer it to
the energy density of the vector. Let's check if inflation
occurs in this model. Using Eqs.(\ref{hamiltonian}) and 
(\ref{evolution:alpha}), equation for acceleration of 
the universe is given by
\begin{eqnarray}
 \ddot{\alpha} + \dot{\alpha}^2 
 = - 2\dot{\sigma}^2 -\frac{\kappa^2}{3} \dot{\phi}^2 
   + \frac{\kappa^2}{3} \left[ V - \frac{p_A^2}{2} f^{-2}
    e^{-4\alpha -4\sigma } \right]  \ .
\end{eqnarray}
We see that the potential energy of the inflaton needs to
be dominant for the inflation to occur.
Now, we assume the energy density of the vector can be  
negligible compared to that of the inflaton for the inflaton dynamics. 
Then, we examine when the anisotropy
is not diluted during inflation. From Eq.(\ref{eq:sigma}),
it is apparent that the fate of anisotropic expansion rate 
$\Sigma \equiv \dot{\sigma}$ depends on the behavior of 
coupling function $f(\phi)$. In the critical 
case $f(\phi ) \propto e^{-2\alpha}$, the energy
density of the vector field as a source term in 
Eq.(\ref{eq:sigma}) remains almost constant during the slow-roll
inflation. Using  slow-roll equations
\begin{eqnarray}
    \dot{\alpha}^2  =  \frac{\kappa ^2}{3}V(\phi), \quad
    3\dot{\alpha}\dot{\phi} = -V'(\phi ) \ ,
    \label{slow1}
\end{eqnarray}
we obtain 
$
    d\alpha / d\phi = \dot{\alpha} /\dot{\phi} 
    = - \kappa ^2 V(\phi) / V'(\phi )  \ . 
$
This can be easily integrated as
$ 
     \alpha =  -\kappa^2 \int V/V' d\phi \ .
$
Here, we have absorbed a constant of integration into the definition of 
$\alpha$. Thus, we obtain
\begin{equation}
f = e^{-2\alpha} =  e^{2\kappa^2 \int \frac{V}{V'} d\phi }  \ .
\label{critical}
\end{equation}
For the polynomial potential $V\propto \phi^n $, we have
$
f =  e^{ \kappa ^2 \phi ^2/n} \ .
$
 Given the critical case (\ref{critical}), 
we can parameterize the coupling function as~\cite{Martin:2007ue}:
\begin{equation}
f = e^{2 c \kappa^2 \int \frac{V}{V'} d\phi } \label{key} \ ,
\end{equation}
where $c$ is a parameter.

Naively, the energy density of the vector field grows during inflation
when $c > 1$, which is the case we want to consider. 
It would not be possible to neglect the vector field 
in this case, and Eq.(\ref{slow1}) would not be 
appropriate for discussing the inflation dynamics anymore.
 Let us see what happens 
if the vector field is not negligible.

\section{Tracking Anisotropic Inflation}
\label{sc:coe}

\begin{figure}[h]
\includegraphics[height=6cm, width=7.5cm]{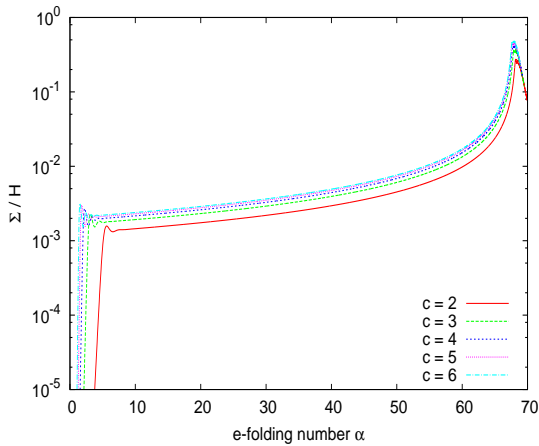}
\caption{
Evolutions of the anisotropy $\Sigma/H$ for various $c$ are shown.
One can see the attractor like behavior of the anisotropy. }
\label{fg:ce-ratio}
\end{figure}

To make the analysis concrete, we consider chaotic inflation with
the potential $V(\phi ) = m^2\phi^2 /2$ ($n=2$). 
For this potential, the coupling function becomes $f(\phi)=e^{c \kappa^2\phi^2 /2}$. 
It is instructive to see what happens by solving
  Eqs.(\ref{hamiltonian})-(\ref{eq:phi}) numerically.
 \begin{figure}[ht]
\includegraphics[height=6cm, width=7.5cm]{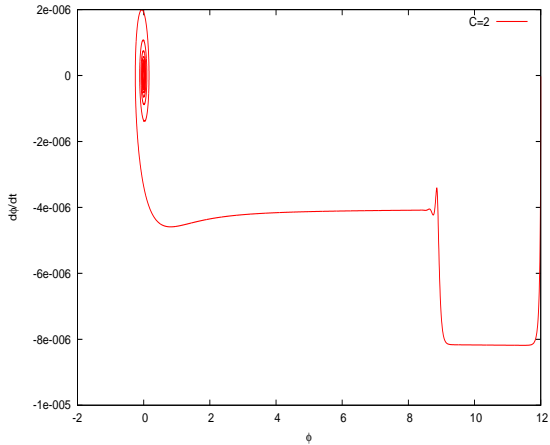}
\caption{Phase flow for $\phi$ is depicted. 
 Here, we took the parameters $c=2$ and 
  $\kappa m=10^{-5} $. We also put initial conditions 
 $\phi_i=12$ and $\dot{\phi}_i=0$.
 There are two different slow-roll phases.
 The transition occurs around $\kappa\phi= 9$.}
\label{fg:phase}
\end{figure}
 In Fig. \ref{fg:phase}, we have shown the phase flow
  in $\phi-\dot{\phi}$ space 
where we can see
  two slow-roll phases, which indicates something different from
 the conventional inflation occurs.
 In Fig.\ref{fg:ce-ratio},
 we have calculated the evolution of the anisotropy 
 $\Sigma/H \equiv \dot{\sigma}/\dot{\alpha}$ for various parameters $c$
 under the initial conditions $\sqrt{c}\kappa\phi_i=17$. 
As expected, all of solutions show a rapid growth of anisotropy
in the first slow-roll phase.
However, the growth of the anisotropy eventually stops at the order of a percent.
Notice that this attractor like behavior is not so sensitive to a parameter $c$.

Now, we will give an analytic explanation of the numerical results
and find a quite remarkable relation between the anisotropy and
a slow-roll parameter of inflation.

As the energy density of the vector 
field should be subdominant during inflation,
we can ignore $\sigma$ in Eqs.(\ref{hamiltonian}), (\ref{evolution:alpha}),
and (\ref{eq:phi}). 
However, in Eq.(\ref{eq:sigma}), 
all terms would be of the same order.
Now, Eqs.(\ref{hamiltonian}) and (\ref{eq:phi}) 
can be written as
\begin{eqnarray}
\dot{\alpha}^2 &=& 
 \frac{\kappa^2}{3}\left[ \frac{1}{2} \dot{\phi}^2
+\frac{1}{2}m^2\phi^2+\frac{1}{2}e^{-c\kappa^2\phi^2-4\alpha } p_{A}^2 
              \right]  \ , \label{h3} \\
\ddot{\phi} &=& -3\dot{\alpha}\dot{\phi} -m^2\phi 
+ c \kappa^2\phi e^{-c\kappa^2\phi^2-4\alpha  }p_{A}^2\label{eq:phi3}      \ .        
\end{eqnarray}
Let's see how the energy density of the vector field works in these equations.
When the effect of the vector field is comparable with that of the inflaton
field as source terms in (\ref{eq:phi3}), we get the relation 
$c\kappa^2 p_A^2 e^{-c\kappa^2 \phi^2 -4\alpha } \sim m^2 $.
If we define the ratio of the energy density of the vector field 
$\rho_A\equiv p_A^2 e^{-c\kappa^2 \phi^2 -4\alpha } /2$ to that of 
the inflaton $\rho_\phi\equiv m^2 \phi^2/2$ as
\begin{equation}
{\cal R} \equiv \frac{\rho_A}{\rho_\phi}
= \frac{p_{A}^2 e^{-c\kappa^2\phi^2-4\alpha}}{m^2\phi^2} \ ,
\label{R}
\end{equation}
we find the ratio becomes
${\cal R} \sim 1 /c\kappa^2 \phi^2$
when the above relation holds.
Since the e-folding number is crudely given by $N\sim \kappa^2 \phi^2$

and the scale observed through CMB corresponds to $N \sim {\cal O} (100)$,
 we have typically  $\kappa \phi \sim {\cal O} (10)$. Hence, the ratio goes
${\cal R} \sim 10^{-2}$. Thus we find that the effect of the
vector filed in (\ref{h3}) is negligible even when it is comparable
with that of the scalar field in (\ref{eq:phi3}).

It turns out that the above situation is not transient one but
an attractor.
Suppose that $\rho_A$ is initially negligible, 
${\cal R}_i \ll 10^{-2} $. In the first slow-roll 
inflationary phase (\ref{slow1}), 
the relation
$e^{-\kappa^2\phi^2} \propto e^{4\alpha} $ 
holds as was shown in (\ref{critical}).
Hence, the ratio ${\cal R}$ varies as
$
{\cal R} \propto e^{4(c-1)\alpha}.
$
As we now consider $c>1$, $\rho_A$ increases rapidly during inflation
and eventually reaches  ${\cal R} \sim 10^{-2}$.
Whereas, when  ${\cal R}$ exceeds $ 10^{-2} $,
the inflaton climbs up the potential due to the effect
of the vector field in (\ref{eq:phi3}), hence $\rho_A$ will decrease rapidly
and go back to the value ${\cal R} \sim 10^{-2}$.
Thus irrespective of initial conditions, $\rho_A$ will track $\rho_{\phi}$.

The above arguments tell us that the inflaton dynamics after tracking is
governed by the modified slow-roll equations
\begin{eqnarray}
 \dot{\alpha}^2 &=& \frac{\kappa^2}{6} m^2 \phi^2   \ , 
\label{h4}\\
3\dot{\alpha} \dot{\phi} &=& 
-m^2\phi+c\kappa^2 \phi p_{A}^2 e^{-c\kappa^2\phi^2-4\alpha  } \ . 
\label{eq:balance}
\end{eqnarray}
We refer to the phase governed by the above equations as the
second inflationary phase, compared to the first one
governed by the equations (\ref{slow1}). 
Using above equations, we can deduce
\begin{equation}
  \phi \frac{d\phi}{d\alpha}  = -\frac{2}{\kappa^2} + \frac{2cp_A^2}{m^2}
                e^{-c\kappa^2 \phi^2 -4 \alpha}  \ . \label{phi:alpha}
\end{equation}
This can be integrated as
$
  e^{-c\kappa^2 \phi^2 -4\alpha} 
  = m^2 (c-1)/ c^2 \kappa^2 p_A^2 \left[1+D e^{-4(c-1)\alpha} \right]^{-1}  ,
$
where $D$ is a constant of integration. This solution rapidly converges 
to 
\begin{eqnarray}
  e^{-c\kappa^2 \phi^2 -4\alpha} 
                = \frac{m^2 (c-1)}{c^2 \kappa^2 p_A^2} \ .
\label{attractor}
\end{eqnarray}
Thus, we found $\rho_A$ becomes constant during the second inflationary 
phase. 
Substituting the result (\ref{attractor}) into 
the modified slow-roll equation (\ref{eq:balance}),
 we obtain the equation for the second inflationary phase
\begin{eqnarray}
  3\dot{\alpha} \dot{\phi} = - \frac{m^2}{c} \phi 
   \label{effective} \ .
\end{eqnarray}
This indicates that $\dot{\phi}$ in the second phase of inflation
 is about $1/c$ times that in the first phase of inflation.
 In Fig. \ref{fg:phase}, 
 we can see the value of $\dot{\phi}$ after the phase transition is about
 a half of that in the first phase, which agrees with the 
analytical estimate for $c=2$.

Now let us consider the anisotropy.
In the second slow-roll phase, Eq.(\ref{eq:sigma}) reads
\begin{eqnarray}
3\dot{\alpha}\dot{\sigma} 
= \frac{\kappa ^2 p_{A}^2}{3}e^{-c\kappa^2\phi^2-4\alpha } 
\label{eq:sigma3} \ .
\end{eqnarray}
where we have assumed 
$\sigma\ll c\kappa^2\phi^2$, $\ddot{\sigma}\ll\dot{\alpha}\dot{\sigma}$.
Using Eqs.(\ref{h4}) and (\ref{eq:sigma3}), the anisotropy turns out 
to be determined by the ratio (\ref{R}) as
\begin{equation}
\frac{\Sigma}{H}  
= \frac{\kappa^2 p_{A}^2 e^{-c\kappa^2\phi^2-4\alpha  }}{9\dot{\alpha}^2}
= \frac{2}{3}{\cal R}(t)  \ .
\label{S/H}
\end{equation}
From Eq.(\ref{attractor}), we can calculate the ratio
\begin{equation}
{\cal R}(t) = \frac{c-1}{c^2\kappa^2\phi^2} 
   \ . \label{ratio}
\end{equation}
Using this relation, we can relate degrees of anisotropy
 to the slow-roll parameter as follows. 
Combining Eqs.(\ref{hamiltonian}) with (\ref{evolution:alpha}),
we obtain
\begin{equation}
\ddot{\alpha} 
=-\frac{\kappa^2}{2}\dot{\phi}^2
   -\frac{\kappa^2}{3}e^{-c\kappa^2\phi^2-4\alpha  }p_{A}^2
\label{eq:ddalpha} \ ,
\end{equation}
where we have used $\dot{\sigma}^2\ll\kappa^2\dot{\phi}^2$
derived from Eqs.(\ref{h4}), (\ref{effective}), (\ref{S/H}) and (\ref{ratio}).
Thus, the slow-roll parameter is given by
\begin{eqnarray}
\epsilon \equiv   -\frac{\ddot{\alpha}}{\dot{\alpha}^2} 
 = \frac{2}{c \kappa^2\phi^2} \ , \label{slow}
\end{eqnarray}
where we used the results (\ref{h4})
, (\ref{attractor}) and (\ref{effective}).
Thus, combining Eqs.(\ref{S/H}),(\ref{ratio}), and (\ref{slow}), 
we reach a main result
\begin{equation}
 \frac{\Sigma}{H} = \frac{1}{3}\frac{c-1}{c} \epsilon \ .
\end{equation}
This remarkable relation shows a quite good agreement
with the numerical results in Fig.\ref{fg:ce-ratio}.

\section{Generality}
\label{sc:generality}

Although the discussion we have made so far is restricted to a specific form
 of potential $V$, we now argue that our finding is the general 
feature of the inflationary scenario in the presence of the 
vector field.
 
Let us consider the general potential $V(\phi)$ for the inflaton.
Then, the coupling function should be of the form (\ref{key}). 
Hence, in the slow-roll phase, the equation for the inflaton (\ref{eq:phi}) 
becomes
\begin{eqnarray}
3\dot{\alpha}\dot{\phi} 
= - V' + 2c\kappa^2 \frac{V}{V'} f^{-2}p_{A}^2 e^{-4\alpha -4\sigma}  \ .
\end{eqnarray}
When $c>1$, the energy density of the vector will soon catch up with
that of the inflaton. At the tracking point,
$\rho_A$ and $\rho_\phi$ tend to be
$
  \rho_A \simeq  \left( V'/V \right)^2 \rho_\phi /4c\kappa^2 \ . 
$
Note that the slow-roll parameter now becomes:
\begin{equation}
\epsilon \equiv -\frac{\ddot{\alpha}}{\dot{\alpha}^2}
   \simeq \frac{1}{2c\kappa^2} \left( \frac{V'}{V}\right)^2 \ .
\end{equation}
Then, again, we can conclude that the anisotropy becomes of the order of the slow-roll parameter:
\begin{equation}
\frac{\Sigma}{H} \simeq \frac{1}{6c\kappa^2}\left( \frac{V'}{V}\right)^2
\simeq \frac{1}{3} \epsilon    \ .
\end{equation}
Thus, we have shown that the anisotropy is universally determined by 
the slow-roll parameter. 
This is reminiscent of non-gaussianity in single 
inflaton models~\cite{Maldacena:2002vr}.

\section{Conclusion}

We have proposed an inflationary scenario with anisotropy. 
Remarkably, we have find that degrees of anisotropy are universally 
determined by the slow-roll parameter of inflation. 
Since the slow-roll parameter is observationally known to be 
of the order of a percent, the anisotropy during inflation
cannot be entirely negligible. 
Indeed, we can expect rich phenomenology as consequences of the
anisotropy during inflation. First of all,
since the rotational invariance is violated, the statistical
anisotropy of CMB temperature fluctuations can be expected
~\cite{Ackerman:2007nb}.
More interestingly, 
tensor perturbations could be induced from curvature perturbations 
through the anisotropy of the background spacetime. One immediate 
consequence is a correlation between 
curvature and tensor perturbations~\cite{Kanno:2008gn}.
This correlation should be detected through the analysis of 
temperature-B-mode correlation in CMB. 
  Moreover, 
because of the anisotropy, there might be linear polarization
in primordial gravitational waves.  
This polarization can be detected either through CMB observations
or direct interferometer observations. 
These predictions can be checked by future observations.
Theoretically, we need more systematic check such as 
quantum loop effects~\cite{Seery:2008ms}.

Finally, let us point out another view of our result.
Our finding of hairy inflation can be regarded as a counter example
 to the cosmic no-hair conjecture. This hair stems from the fact
 that the inflation is not exactly deSitter expansion. In fact, degrees of
 anisotropy is determined by the slow-roll parameter.
 In a sense, this is the origin of the universality of
 a percent level of vector hair.

\begin{acknowledgements}
JS is supported by the Japan-U.K. Research Cooperative Program, 
Grant-in-Aid for  Scientific Research Fund of the Ministry of 
Education, Science and Culture of Japan No.18540262 and No.17340075. 
\end{acknowledgements}

\end{document}